# Structural and magnetic properties of lightly doped M-type hexaferrites

Sami H. Mahmood[1a], Ahmad M. Awadallah[1b], Ibrahim Bsoul[2c], Yazan Maswadeh[3d]

[1] Physics Department, The University of Jordan, Amman 11942, Jordan
[2] Physics Department, Al al-Bayt University, Mafraq 13040, Jordan
[3] Physics Department, Central Michigan University, Mount Pleasant 48859, MI, USA

[a]s.mahmood@ju.edu.jo (Corresponding author), [b]ahmadmoh@yahoo.com

[c]ibrahimbsoul@yahoo.com, [d]nawabra251@gmail.com

**ABSTRACT**

Vanadium substituted SrM hexaferrites ($SrFe_{12-x}V_xO_{19}$ with $x = 0.2, 0.4$) and Eu-substituted BaM hexaferrite ($Ba_{0.8}Eu_{0.2}Fe_{12}O_{19}$) were prepared by high energy ball milling and sintering at 1200° C. X-ray diffraction measurements revealed that the V-substituted SrM samples exhibited phase separation resulting in the coexistence of the pure SrM magnetic phase with nonmagnetic $Sr_3(VO_4)_2$ vanadate and $\alpha$-$Fe_2O_3$ iron oxide phase. Also, the Eu-substituted BaM hexaferrite revealed the formation of the pure BaM phase coexisting with $\alpha$-$Fe_2O_3$ secondary phase, and Eu-garnet minor phase. Although the magnetic properties of the samples deteriorated with respect to pure hexaferrite properties, the magnetic parameters of the substituted samples were found to be of potential importance for practical applications. Further, the results of the study suggest methods for the preparation of high quality SrM hexaferrites, and hexaferrite/garnet composites.

**Section Headings**

**1. Introduction**

**2. Experimental**

1.     INTRODUCTION

Magnetic materials play a very important role in our life; magnets have been used in a wide range of industrial and technological applications such as data processing, electronic devices, telecommunication, automobile industry, loudspeakers, instrumentation, power generation, motors for a wide spectrum of applications, and so on [1, 2]. Hexaferrites belong to a special kind of magnetic oxides which demonstrated potential for various applications including permanent magnet and microwave applications. Three decades after their discovery in the 1950s, the annual production of M-type hexaferrite dominated the world market of permanent magnets due to cost effectiveness, easy production, corrosion resistance, and low eddy current losses [3]. The production and characterization of hexaferrites serves in both fields of scientific fundamental research and technological applications as demonstrated by the growth of market demand, and the exponential increase of the annual number of publications and registered patents [4]. For example, more than 50% of the magnetic materials produced annually around the world consist of M-type barium hexaferrite (BaM) [5].

Soft ferrites are usually used in transformers [6] and magnetic recording heads [7], multilayer chip inductors (MLCIs) [8, 9], soft magnets in electronic components such as power supplies [9-11], and in microwave high frequency devices [5, 12]. However, hard hexaferrites such as BaM have large magnetization and high coercive field, and is thus suitable for use in high-density magnetic recording media [13-19], permanent magnets [5, 10, 20], magneto-optics [21], and microwave absorption devices in GHz range [11, 15, 22]. Electromagnetic interference (EMI) could lead to severe interruption of the functioning of electronically controlled systems due to the electronic pollution produced by gigahertz (GHz) electronic telecommunication systems. For this reason, interest in producing and modifying electromagnetic wave absorbers and filters grew in the last few years. The M-type hexaferrites with planar magnetic anisotropy have better properties as electromagnetic wave absorbers in the GHz range compared to conventional Ni-Zn and Mn-Zn cubic ferrites [23, 24]. Cubic ferrites are widely used in various inductive devices with working frequencies below 100 MHz, but the problem with those ferrites arise when the working frequencies exceed 100 MHz [9, 25]. On the contrary, M-type hexaferrites have large dielectric and magnetic losses in the microwave frequency band [25]. In addition to their low production cost [26], low density, simple and easy fabrication process, and excellent chemical and thermal stability [11, 13, 15, 27], M-type hexaferrites have gained special attention in the field of materials research due to their interesting and attractive properties. These include, and are not limited to large saturation magnetization [15, 28], high coercive field with strong anisotropy along the *c*-axis [29, 30], high Curie temperature [11], very low electrical conductivity [13, 23], mechanical hardness [13], corrosion resistivity [11, 31], high microwave magnetic loss [22], moderate permittivity [32], and chemical compatibility with biological tissues [13, 33].

Based on their structures and chemical compositions, barium-based hexaferrites are classified into six main types [34]. With the spinel formula (S) referring to $Me_2Fe_4O_8$ these types are:

- M-type: $[BaFe_{12}O_{19}] = [(BaO).6(Fe_2O_3)] = M$
- W-type: $[BaMe_2Fe_{16}O_{27}] = [(BaO)\cdot2(MeO)\cdot8(Fe_2O_3)] = M+S$
- X-type: $[Ba_2Me_2Fe_{28}O_{46}] = [2(BaO)\cdot2(MeO)\cdot14(Fe_2O_3)] = 2M+S = M+W$
- Y-type: $[Ba_2Me_2Fe_{12}O_{22}] = [2(BaO)\cdot2(MeO)\cdot6(Fe_2O_3)] = Y$
- Z-type: $[Ba_3Me_2Fe_{24}O_{41}] = [3(BaO)\cdot2(MeO)\cdot12(Fe_2O_3)] = M+Y$
- U-type: $[Ba_4Me_2Fe_{36}O_{60}] = [4(BaO)\cdot2(MeO)\cdot18(Fe_2O_3)] = 2M+Y$

In all of these ferrites, Ba can be replaced by Sr, Pb, or other elements [35]. The structure of BaM hexaferrite is hexagonal with $P6_3/mmc$ space group and lattice parameters $a = b = 5.89$Å, $c = 23.17$Å, $\alpha = \beta = 90°, \gamma = 120°$ [5]. The unit cell of BaM with magnetoplumbite-type structure includes two $BaFe_{12}O_{19}$ molecules as seen in Fig. 1 [36], each molecule arranged in two structural blocks: the hexagonal R block and the spinel S block [4, 36].

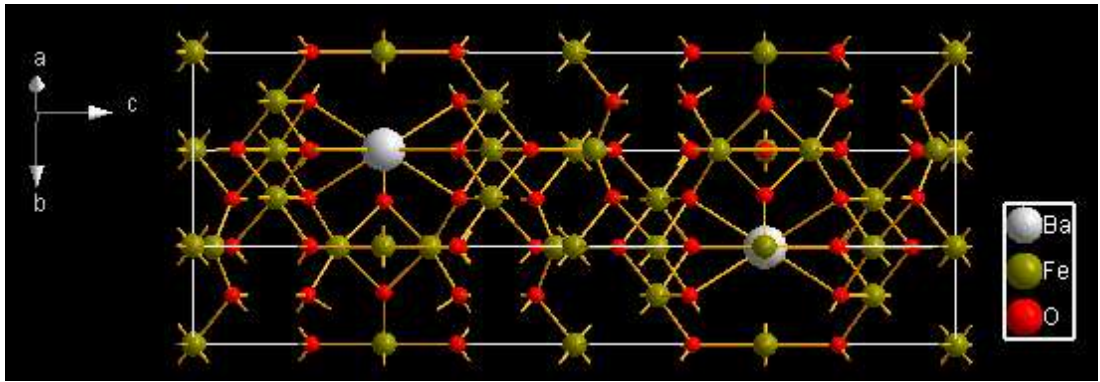

Fig. 1: BaM unit cell.

The unit is composed of the stacking sequence SRS*R*, where S* and R* blocks are S and R blocks rotated by 180º about the hexagonal *c*-axis. Within the S block, there are three interstitial sites between the oxygen layers which are occupied by the small metallic ions like $Fe^{3+}$ ions [32]. One of these is the six-coordinated octahedral (2a) site occupied by spin-up iron ion, while the other two are four-coordinated tetrahedral ($4f_1$) sites occupied by spin-down iron ions. Also, within the R block there are three interstitial sites between oxygen layers which are occupied by the small metallic ions, two six-fold coordinated octahedral ($4f_2$) sites occupied by spin-down iron ions, and one trigonal bi-pyramidal five-fold local symmetry (2b) site occupied by spin-up iron cation. Three interstitial octahedral (12k) sites are also available within each R-S interface layer, which are occupied by spin-up metallic ions.

Table. 1 shows the spin orientation of the $Fe^{3+}$ at the different sites in the unit cell. According to Gorter's model of a system of collinear spin structure [37], the net magnetic moment per molecule is given by the sum of the magnetic moments of magnetic ions in the molecule. For BaM molecule ($BaFe_{12}O_{19}$), the S block contributes 2 spin down and 1 spin up moments, the R block contributes 2 spin down and 1 spin up moments, and both S-R interfaces contribute 6 spin up moments. Accordingly, the net magnetic moment per formula ($BaFe_{12}O_{19}$) can be calculated as follows:

$$m = (1\mu_{2a} - 2\mu_{4f1})_S + (6\mu_{12k})_{S-R} + (1\mu_{2b} - 2\mu_{4f2})_R = 4\mu \qquad (1)$$

With a moment of $\mu = 5\mu_B$ per $Fe^{3+}$ ion, the magnetic moment of $BaFe_{12}O_{19}$ is $20\mu_B$ per molecule. This corresponds to magnetization value of about 100 emu/g. This value was confirmed by measurements of the magnetization of SrM at low temperature where the thermal effects are minimized [38]. At room temperature, the magnetization for a typical BaM material was reduced to about 72 emu/g by thermal effects [5]. Also, typical coercive fields of 3 – 5 kOe were observe for BaM materials [5]. However, large variations in saturation magnetizations and coercive fields were reported, and were attributed to variations in particle size and morphology, as well as to different preparation methods [35]. Also, the substitution of Ba and/or Fe by other metals was found to influence the magnetization and coercivity significantly [35].

The effect of rare-earth ions substitution for the divalent metal ions was found to result in significant variations of the magnetic properties of M-type hexaferrites. Specifically, the substitution of Ba by Eu in $Ba_{0.75}Eu_{0.25}Fe_{12}O_{19}$ prepared sol–gel method was found to result in an increase in coercivity and a decrease in saturation magnetization [39]. Also, 10% substitution of Ba by La in BaM prepared by a reverse micro-emulsion route was found to lead to high magnetic properties, and enhancement of the microwave absorption properties for potential microwave absorption applications [40]. In addition, it was found that the saturation magnetization of $Sr_{1-x}La_xFe_{12-x}Co_xO_{19}$ improved upon increasing $x$ up to 0.2, and the coercivity increased with increasing $x$ up to 0.3 [41]. Further, La and Pr substitution for Ba in BaM powders prepared by auto-combustion was found to result in an increase in saturation magnetization and coercivity of the ferrite [42].

This work is a continuation of the recently reported study on the effect of vanadium substitution for iron on the structural and magnetic properties of BaM hexaferrites [18]. In addition, the study of the effect of RE substation was motivated by the reported influence of RE elements on the properties of M-type hexaferrites.

*Table 1: The spin orientation with its corresponding site for the unit cell.*

| Block | Sites | symmetry | Cations/site | Spin Orientation | Total spin magnetic moment |
|---|---|---|---|---|---|
| S | 2a | Octahedral | 1 | Spin up ( ↑ ) | 1 ( ↓ ) |
|   | 4f1 | Tetrahedral | 2 | Spin down ( ↓ ) |   |
| S-R | 12k | Octahedral | 3 | Spin up ( ↑ ) | 3 ( ↑ ) |
| R | 2b | Bi-pyramidal | 1 | Spin up ( ↑ ) | 1 ( ↓ ) |
|   | 4f2 | Octahedral | 2 | Spin down ( ↓ ) |   |
| R-S* | 12k | Octahedral | 3 | Spin up ( ↑ ) | 3 ( ↑ ) |
| S* | 2a | Octahedral | 1 | Spin up ( ↑ ) | 1 ( ↓ ) |
|   | 4f1 | Tetrahedral | 2 | Spin down ( ↓ ) |   |
| S*-R* | 12k | Octahedral | 3 | Spin up ( ↑ ) | 3 ( ↑ ) |
| R* | 2b | Bi-pyramidal | 1 | Spin up ( ↑ ) | 1 ( ↓ ) |
|   | 4f2 | Octahedral | 2 | Spin down ( ↓ ) |   |
| R*-S | 12k | Octahedral | 3 | Spin up ( ↑ ) | 3 ( ↑ ) |

(Formula unit: S through R-S*; Unit cell: all rows)

## 2. EXPERIMENTAL

Powder precursors of M-type hexaferrites were prepared by wet grinding using a high energy ball mill (Fritsch Pulverisette-7). $SrFe_{12-x}V_xO_{19}$ (with $x$ = 0.2, 0.4) was prepared with from high purity (> 98%) $SrCO_3$, $Fe_2O_3$, and $V_2O_3$. Required amounts of these materials were weighed carefully, and 5 g of the starting powder was transferred to each of the two zirconia cups of the ball-mill with 8 mL of acetone. Seven zirconia balls of approximately 10 g each were used for grinding, so that the powder-to- ball mass ratio was 1:14. The grinding was maintained for 16 hours with 250-rpm rotational speed, and the grinding was carried out in intervals of 10 min. each separated by 5 min. pause periods to avoid overheating. The dried powder was collected, and about 0.8 g discs (1.25 cm in diameter and ~ 1 mm thick) were prepared by pressing under a force of 4 tons. The discs were then sintered at a temperature of 1200º C for 2 in a zirconium oxide crucible. Also, $Ba_{0.8}Eu_{0.2}Fe_{12}O_{19}$ sample was prepared according to the same procedure, except that the sintering process was carried out in an alumina crucible.

Structural details of the samples were obtained by analyzing the x-ray diffraction patterns obtained using XRD 7000-Shimadzu diffractometer with Cu-Kα radiation ($\lambda = 1.5405$ Å). The samples were scanned over the angular range $20° < 2\theta < 70°$ with 0.01° scanning step and speed of 0.5 deg/min. The grain morphology and grain size of the samples were examined by Scanning Electron Microscope (SEM) system (FEI-Inspect F50/FEG). The magnetic properties of the samples were investigated using a vibrating sample magnetometer (VSM Micro Mag 3900, Princeton Measurements Corporation). Samples for VSM measurements were prepared by cutting small pieces from the sintered discs and gently polishing to the desired needle-shape, to reduce the shape anisotropy effect. The magnetic data were carried out at room temperature with an applied magnetic field up to 10 kOe.

## 3. RESULTS AND ANALYSIS
## 3.1. STRUCTURAL RESULTS

The XRD patterns of the samples $SrFe_{12-x}V_xO_{19}$ ($x = 0.2, 0.4$) are shown in Fig. 2. Rietveld refinement of the patterns indicated each sample is composed of four structural phases: $SrFe_{12}O_{19}$, $\alpha$-$Fe_2O_3$, $Sr_3(VO_4)_2$, and $ZrO_2$. The fitting was reliable as indicated by the low goodness of fit ($\chi^2 = 1.18$ for the sample with $x = 0.2$, and 1.25 for the sample with $x = 0.4$). The appearance of the zirconia phase in the samples is an indication that the samples were contaminated as a result of the high sintering temperature in a zirconium oxide crucible, and this result is consistent with the appearance of this phase in similarly heat treated $Cr_2Y$ samples [34]. The appearance of the $\alpha$-$Fe_2O_3$ phase, on the other hand, resulted from the consumption of a fraction of $Sr^{2+}$ ions commensurate with the stoichiometry of the $Sr_3(VO_4)_2$ vanadate phase, leaving the Fe:Sr molar ratio in the sample higher than the stoichiometric ratio of SrM phase. The reliability factors of Rietveld refinement and the phase proportions (in wt. %) are shown in Table 2.

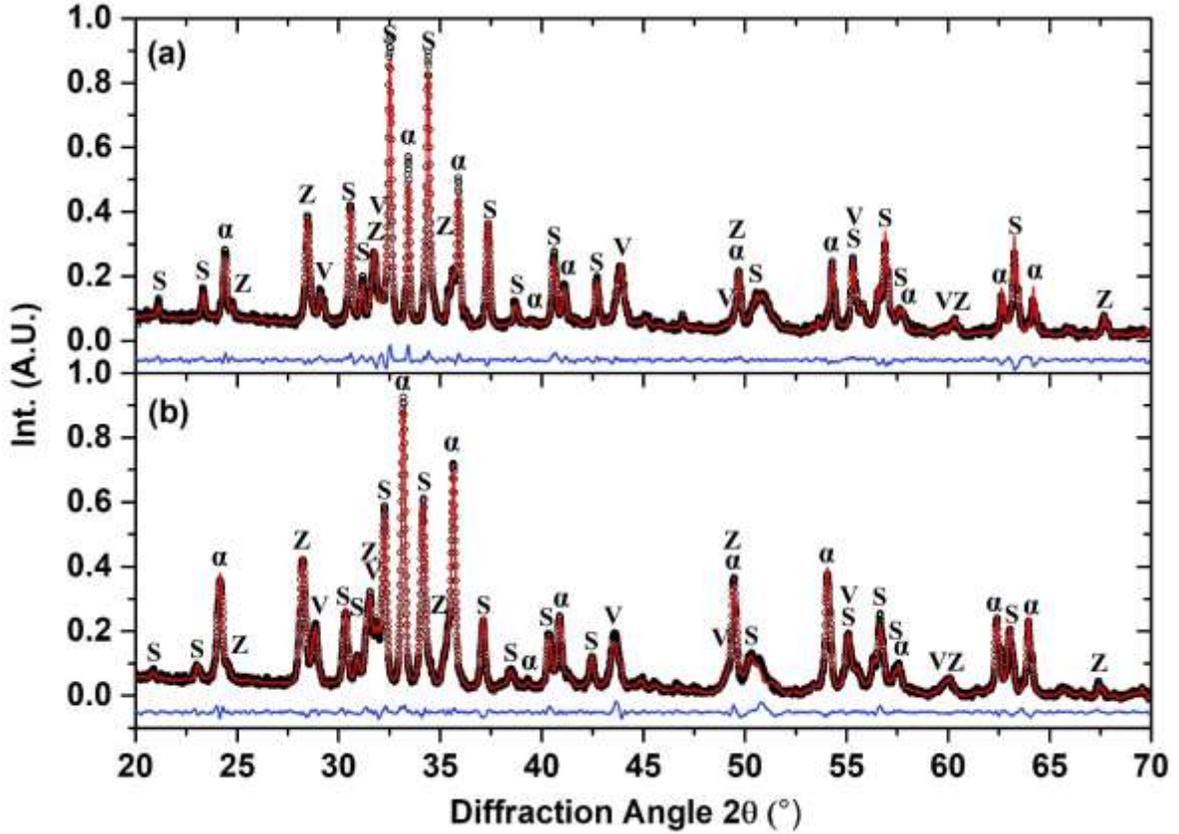

Fig. 2: XRD patterns for the samples $SrFe_{12-x}V_xO_{19}$: (a) $x = 0.2$, (b) $x = 0.4$. Phase contribution in the XRD peaks was labeled as S: M-type Strontium hexaferrite ($SrFe_{12}O_{19}$), α : Hematite ($\alpha$-$Fe_2O_3$), V : Strontium Vanadium Oxide ($Sr_3(VO_4)_2$), Z: Zirconia ($ZrO_2$).

*Table 2: Rietveld fitting reliability factors and fractions of phases (in wt. %) in V-substituted SrM hexaferrite samples.*

| Phase formula | $R_B$ | | $R_F$ | | JCPDS File # | Wt. % | |
|---|---|---|---|---|---|---|---|
| | $x = 0.2$ | $x = 0.4$ | $x = 0.2$ | $x = 0.4$ | | $x = 0.2$ | $x = 0.4$ |
| $SrFe_{12}O_{19}$ | 4.38 | 3.22 | 3.11 | 2.51 | 00-024-1207 | 62.38 | 39.68 |
| $Sr_3(VO_4)_2$ | 8.28 | 4.77 | 6.86 | 2.65 | 00-029-1318 | 3.17 | 3.99 |
| $\alpha$-$Fe_2O_3$ | 4.71 | 1.98 | 3.23 | 1.37 | 00-013-0534 | 23.58 | 43.81 |
| $ZrO_2$ | 3.74 | 2.88 | 3.03 | 1.88 | 00-013-0307 | 10.87 | 12.52 |

The results of the refinement of the XRD pattern for the sample with $x = 0.2$ indicated that the SrM hexaferrite was the major phase, and that the iron oxide and zirconium oxide phases coexisted with appreciable weight fractions of the sample. The $Sr_3(VO_4)_2$ phase appeared with a relatively low weight fraction due to the small amount vanadium in the sample. The effect of the vanadium substitution, however, is more effective in producing α-$Fe_2O_3$ phase, since the formation of one mole of vanadate (requiring 3 moles of Sr) should be accompanied with a surplus of 36 moles of Fe (18 moles of α-$Fe_2O_3$), since the Sr:Fe in the SrM is 1:12. The refinement results, therefore, indicated that the weight fraction of the α-$Fe_2O_3$ phase increased significantly in the sample with $x = 0.4$, which is due to the higher consumption of Sr by the vanadate phase, and the consequent depletion of Sr available for the formation of the SrM phase.

The crystallite sizes of the different samples were calculated using the Stokes-Wilson relation [43]:

$$D = \frac{\lambda}{\beta \cos \theta} \qquad (2)$$

Here $D$ is the crystallite size, $\lambda$ is the wavelength of radiation (1.5406 Å), $\beta$ is the integral breadth, and $\theta$ is the peak position. The integral breadth was corrected for instrumental broadening using Si standard sample. The crystallite size was determined from the (110) peak at $2\theta = 30.3°$, the (107) peak at $2\theta = 32.2°$, and the (114) peak at $2\theta = 34.1°$. Table 3 shows the crystallite size along the different crystallographic directions. These results indicate that the average crystallite size along the basal plane (the <110> direction) is larger than that along the c-axis, suggesting that the crystallites grow in platelet-like shapes. Further, higher concentrations of V-doping resulted in a significant reduction of the crystallite size, indicating poor crystallinity. This effect could be due to the hindrance of the growth of the SrM crystallite by the crystallization of the secondary phases in the sample.

Table 3: Crystallite size along different crystallographic directions for $SrFe_{12-x}V_xO_{19}$ hexaferrites.

| x | D (nm) | | |
|---|---|---|---|
|  | (1 1 0) | (1 0 7) | (1 1 4) |
| 0.2 | 141 | 93 | 114 |
| 0.4 | 88 | 52 | 35 |

Fig. 3 shows the XRD pattern for the sample $Ba_{0.8}Eu_{0.2}Fe_{12}O_{19}$. Rietveld fitting of the diffraction pattern of this sample indicated that the experimental pattern is well fitted ($\chi^2 = 1.13$) with a superposition of the patterns of $BaFe_{12}O_{19}$, α-$Fe_2O_3$, and the garnet phase $Eu_3Fe_5O_{12}$. The reliability factors and the fractions (in wt. %) of the phases derived from Rietveld refinement of the diffraction pattern are listed in Table 4.

Table 4: Fitting reliability factors and fractions of phases in the sample $Ba_{0.8}Eu_{0.2}Fe_{12}O_{19}$, determined by Rietveld refinement of the diffraction pattern.

| Phase | $R_B$ | $R_F$ | JCPDS File # | Wt.% |
|---|---|---|---|---|
| $BaFe_{12}O_{19}$ | 2.47 | 2.52 | 00-027-1029 | 79.9 |
| $\alpha\text{-}Fe_2O_3$ | 3.26 | 3.13 | 00-013-0534 | 17.5 |
| $Eu_3Fe_5O_{12}$ | 5.39 | 4.53 | 00-023-1069 | 2.60 |

These results indicated that the Eu-garnet phase is present with a very low fraction, compatible with the amount of Eu in the sample. According to the refinement results, the sample was phase separated according to the reaction scheme:

$Ba_{0.8}Eu_{0.2}Fe_{12}O_{19} \rightarrow (0.8)BaFe_{12}O_{19} + (1.03)Fe_2O_3 + (0.067)Eu_3Fe_5O_{12}$

The crystallite size was calculated using Eq. 2. The average crystallite size along the <110> direction was found to be 34 nm, which is essentially the same as the value along the <107> direction. The crystallite size in the <114> direction, however, was found to be 50 nm. This indicates that the crystallites of BaM seem to grow into cuboidal shapes rather than platelet shapes. The relatively small crystallite size in this sample could also be associated with the hindrance of the growth of BaM crystallites by the crystallization of neighboring secondary phases.

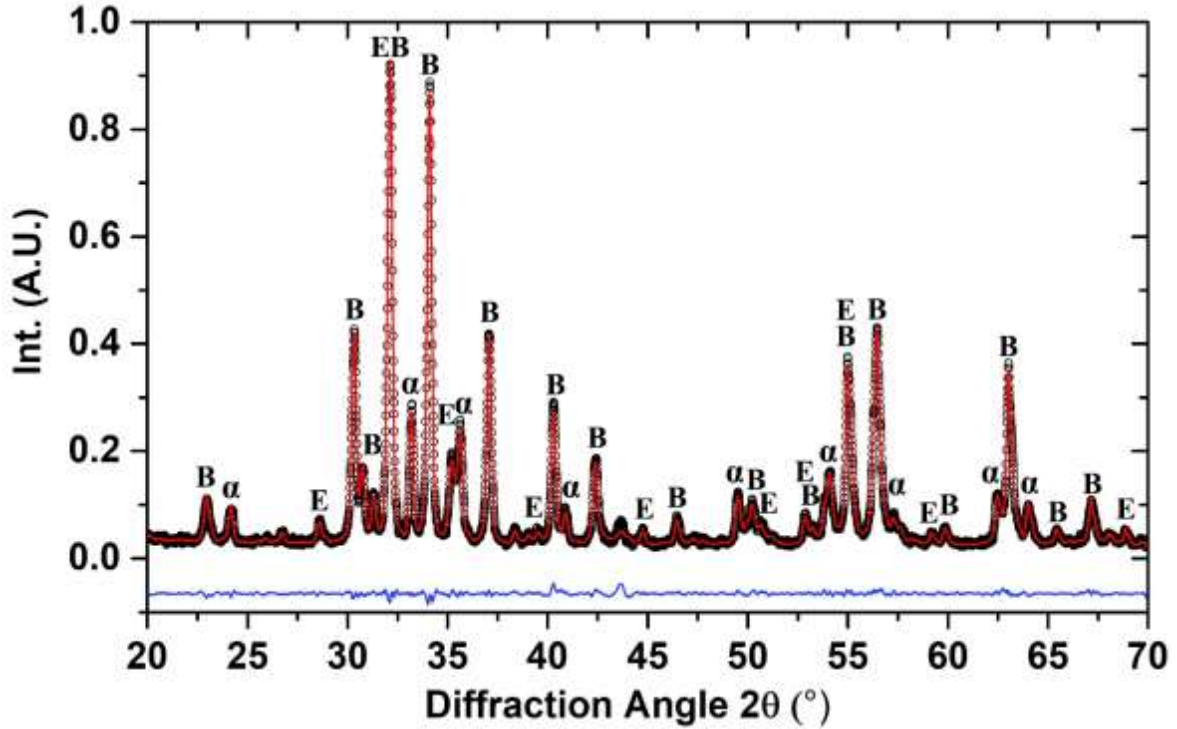

Fig. 3: XRD pattern for the samples $Ba_{0.8}Eu_{0.2}Fe_{12}O_{19}$. Phase contribution in the XRD peaks was labeled as B: M-type barium hexaferrite ($BaFe_{12}O_{19}$), α : Hematite (α-$Fe_2O_3$), E: Europium Iron Oxide ($Eu_3Fe_5O_{12}$)

### 3.2. SEM MEASUREMENTS

The SEM images in Fig. 5 show agglomerations of particles with generally wide particle size distributions for all samples. The particle size for the sample $SrFe_{11.8}V_{0.2}O_{19}$ ranged from about 400 nm to 3.0 µm, with appreciable fraction of the particles characterized by particle sizes below 1 µm. On the other hand, SEM image for the sample $SrFe_{11.6}V_{0.4}O_{19}$ shows particles ranging from 400 nm up to 3.6 µm in size, and general particle growth manifested by the reduction of the fraction of smaller particles in this sample. For the BaM sample doped by Eu, the SEM image shows particles with generally smaller sizes compared with the V-doped SrM system. The particle size for this sample ranged from 200 nm to 1300 nm, with the majority of the particles characterized by particle sizes below 1 µm.

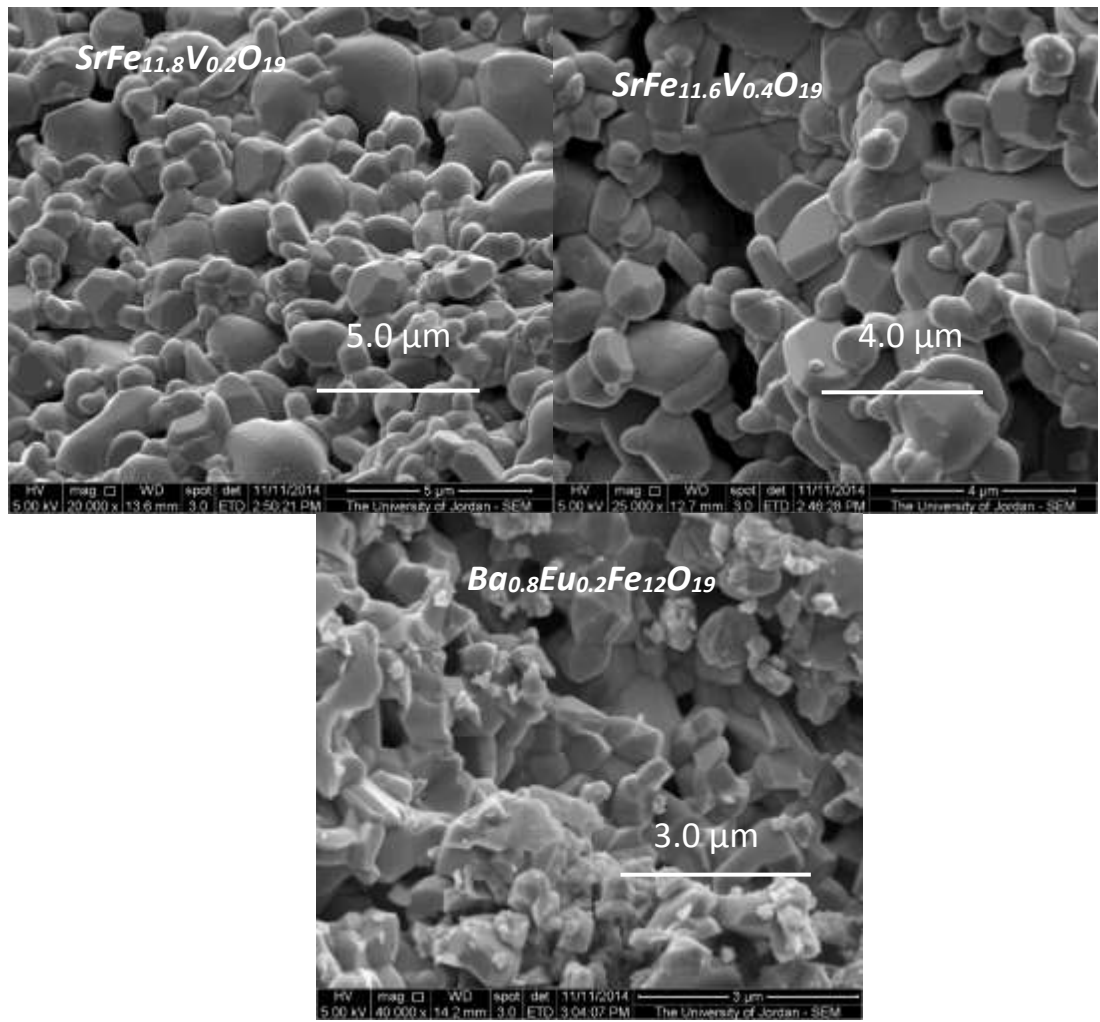

Fig. 4: SEM images for SrFe$_{12-x}$V$_x$O$_{19}$ ($x$ = 0.2, 0.4) and Ba$_{1-x}$Eu$_x$Fe$_{12}$O$_{19}$ ($x$ = 0.2) samples.

### 3.3. MAGNETIC MEASUREMENTS

The room temperature magnetic measurements were carried out using VSM in an applied field up to 10 kOe. The HL for all samples are shown in Fig. 5. The remanence magnetization and the coercive fields were determined directly from the loops. The hysteresis loops showed a behavior characteristic of hard magnetic material, indicated by the absence of magnetic saturation at the upper limit of the applied field. Accordingly, the law of approach to saturation was used to obtain the saturation magnetization. A plot of $M_s$ vs. $1/H^2$ in the high field region (8.6 kOe < $H$ < 10 kOe) for the different samples gave perfect straight lines as demonstrated by Fig. 6. From these straight lines, the saturation magnetization $M_s$, as well as the anisotropy field $H_a$ were determined following the procedure explained elsewhere [16-19, 30]. The magnetic parameters of the samples are listed in Table 5.

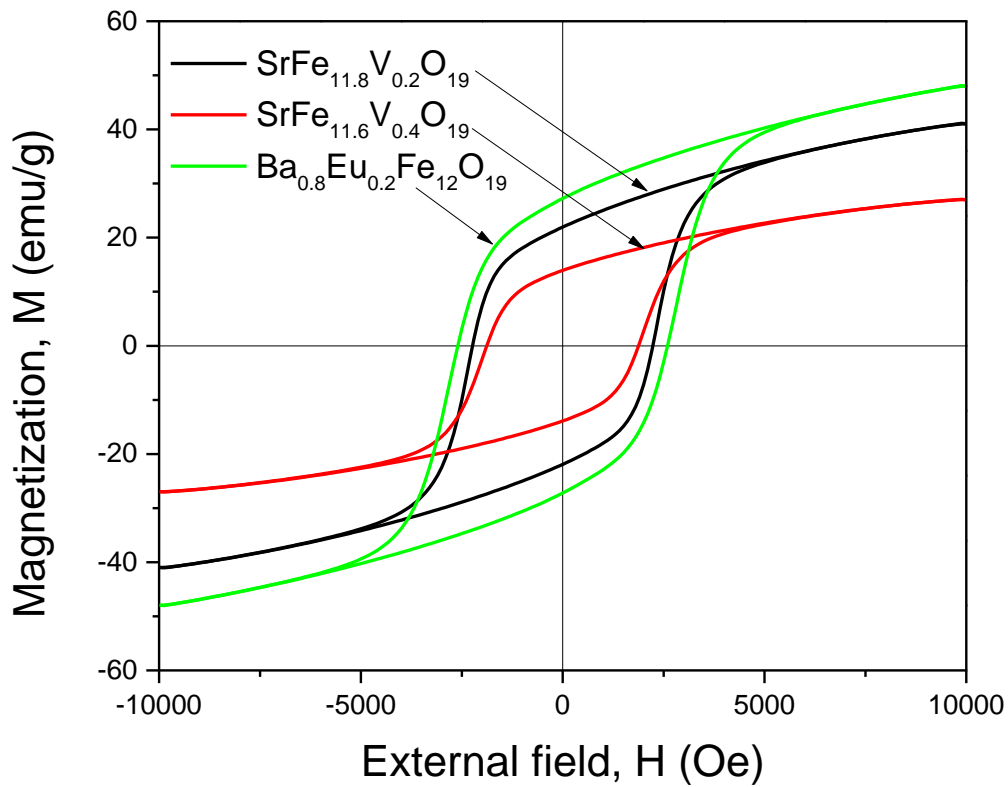

Fig. 5: Hysteresis loops for the samples $SrFe_{12-x}V_xO_{19}$ ($x$ = 0.2, 0.4) and $Ba_{0.8}Eu_{0.2}Fe_{12}O_{19}$.

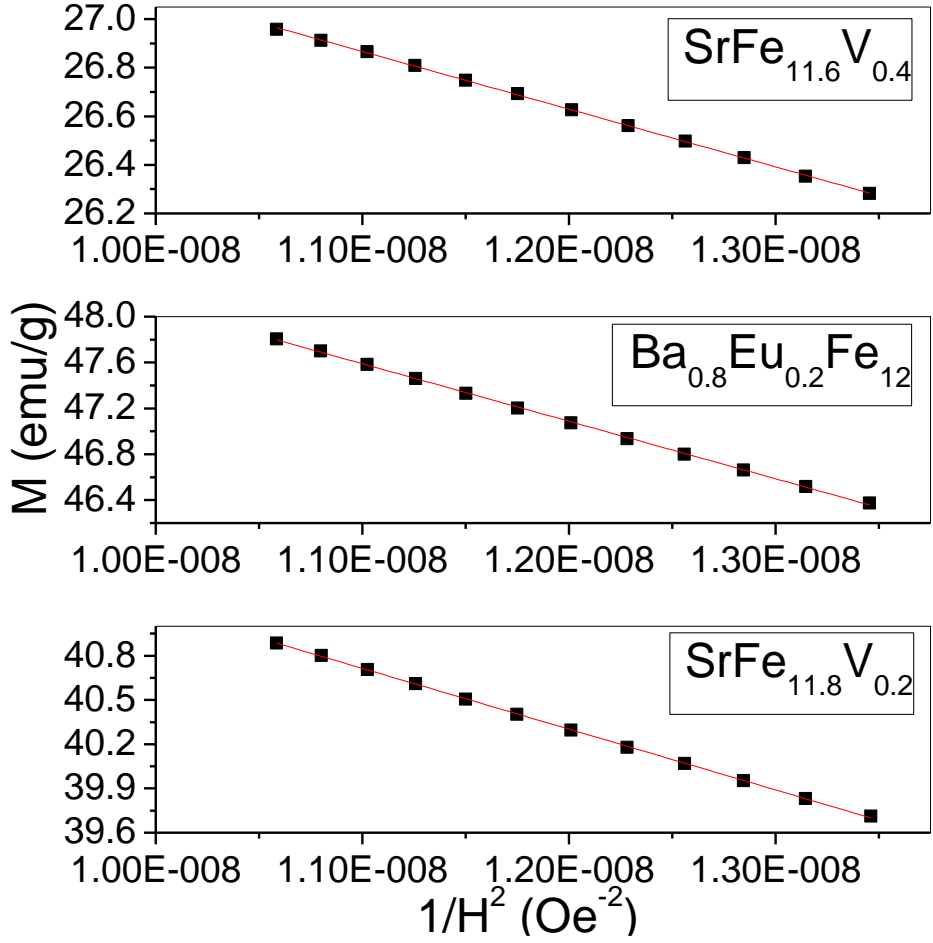

Fig. 6: Magnetization versus $1/H^2$ for the samples $SrFe_{12-x}V_xO_{19}$ ($x$ = 0.2, 0.4) and $Ba_{0.8}Eu_{0.2}Fe_{12}O_{19}$.

Table 5: Saturation magnetization ($M_s$), remanence ($M_r$), squareness ($M_{rs} = M_r/M_s$), coercive field ($H_c$), and anisotropy field ($H_a$) for the samples $SrFe_{12-x}V_xO_{19}$ ($x$ = 0.2, 0.4) and $Ba_{0.8}Eu_{0.2}Fe_{12}O_{19}$.

| Substitution | $M_s$ (emu/g) | $M_r$ (emu/g) | $M_{rs}$ (emu/g) | $H_c$ (kOe) | $H_a$ (kOe) |
|---|---|---|---|---|---|
| SrM-0.2 V | 45.2 | 21.9 | 0.49 | 2.22 | 11.6 |
| SrM-0.4 V | 29.4 | 13.9 | 0.47 | 1.87 | 10.9 |
| BaM-0.2 Eu | 53.1 | 27.2 | 0.51 | 2.59 | 11.9 |

The magnetic data revealed a reduction of the saturation magnetization of all samples with respect to previously reported values of 70 emu/g or higher for pure BaM and SrM compounds [5, 35, 44, 45]. This reduction is a consequence of the formation of nonmagnetic phase in the samples. The value of $M_s$ for the V-substituted SrM sample with $x = 0.2$ is 45.2 emu/g, which is lower than the value of 56.8 emu/g for BaM with 0.2 V-substitution for Fe [18]. Considering that the wt. % of the hexaferrite magnetic phase in this sample is 62.38%, the saturation magnetization of the hexaferrite phase (normalized to its weight fraction) would be 72.5 emu/g. Also, the value of 29.4 emu/g for the sample with $x = 0.4$ is lower than that of 38 emu/g for the similarly substituted BaM [18]. However, the wt. % of the hexaferrite phase in this sample is only 39.68 %, implying that the saturation magnetization of the hexaferrite phase is 74.1 emu/g. The normalized saturation magnetizations are in agreement with the highest values reported for pure SrM hexaferrites [5]. Accordingly, this synthesis route can be adopted to produce high quality SrM magnets by following the procedure recently reported for the production of BaM hexaferrites [18, 46], namely, adding an extra amount of Sr enough to eliminate the occurrence of α-$Fe_2O_3$ phase, avoiding sintering the powder mixture in the zirconium oxide crucible, and washing the sintered powder with HCl to etch out the vanadate nonmagnetic phase. The coercivities of these samples of 2.22 kOe and 1.87 kOe, respectively, are lower than those of similarly substituted BaM [18]. The reduced coercivity could be due to larger particle size in substituted SrM samples. The magnetic properties of these samples could be of potential importance for high density magnetic recording applications [47].

The sample $Ba_{0.8}Eu_{0.2}Fe_{12}O_{19}$ also exhibited a saturation magnetization of 53.1 emu/g, and coercivity of 2.59 kOe. The saturation magnetization is only slightly higher than that reported by Khademi et al. [39] for similarly doped BaM samples prepared by sol-gel method, who found these materials of potential importance for microwave applications in the GHz region. Considering the weight fraction of the BaM phase of 79.9%, and ignoring the small magnetic contribution of the 2.6 wt. % of the garnet phase, the saturation magnetization of the BaM phase, normalized to its weight fraction in the sample, would be 66.5 emu/g. This value is close to the highest reported experimental values for pure BaM hexaferrites [35]. The results of the present work suggests a method for the preparation of hexaferrite/garnet composites with desired composition. This would involve the choice of the Eu amount to provide the required fraction of the garnet phase, adding the amount of Fe necessary for that phase, and then adding stoichiometric amounts of Ba and Fe necessary for the BaM phase. The product composite consisting of the self-biased BaM phase [32] together with the soft garnets could therefore be tuned for microwave applications in the desired microwave band.

The magnetocrystalline anisotropy field ($H_a$) for all investigated samples remained close to the reported value of ~ 12 kOe for pure and substituted BaM [30, 48]. This is due to the fact that the anisotropy field is characteristic of the pure magnetic phase, and is not influenced by the coexisting nonmagnetic phases. Further, the values of $M_{rs}$ ~ 0.5 are consistent with the existence of particles consisting of single magnetic domains of size ≤ 1 µm [5].

## 4. CONCLUSIONS

**The partial substitution of Fe by V in SrM hexaferrite resulted in phase separation and the consequent appearance of vanadate and α-$Fe_2O_3$ nonmagnetic phases coexisting with the magnetic hexaferrite phase. This phase separation resulted in a reduction of the saturation magnetization and coercivity with respect to pure SrM phase. The sample with 0.2 V-substitution exhibited magnetic properties which could be of importance for high density magnetic recording applications. On the other hand, partial substitution of Ba by Eu in BaM hexaferrite was not successfully achieved. Instead, the Eu was involved in forming Eu-garnet phase, and the remaining precursor powders formed pure BaM and α-$Fe_2O_3$. As a consequence, the saturation magnetization of the sample decreased. However, the saturation magnetization of the sample remained high enough for practical applications, and the reduction of the coercivity makes this sample of potential importance for high density magnetic recording.**

ACKNOWLEDGEMENTS

This work was supported by the Deanship of scientific Research and Quality Assurance at The University of Jordan.